# `IASelect`: Finding Best-fit Agent Practices in Industrial CPS Using Graph Databases

Chandan Sharma[1], Roopak Sinha[2], Paulo Leitão[3]
[1&2]IT & Software Engineering, Auckland University of Technology, New Zealand
email: chandan.sharma, roopak.sinha@aut.ac.nz
[3]Research Centre in Digitalization and Intelligent Robotics (CeDRI), Instituto Politécnico de Bragança,
Campus de Santa Apolónia, 5300-253 Bragança, Portugal
email: pleitao@ipb.pt

**Abstract**

*The ongoing fourth Industrial Revolution depends mainly on robust Industrial Cyber-Physical Systems (ICPS). ICPS includes computing (software and hardware) abilities to control complex physical processes in distributed industrial environments. Industrial agents, originating from the well-established multi-agent systems field, provide complex and cooperative control mechanisms at the software level, allowing us to develop larger and more feature-rich ICPS. The IEEE P2660.1 standardisation project, "Recommended Practices on Industrial Agents: Integration of Software Agents and Low Level Automation Functions" focuses on identifying Industrial Agent practices that can benefit ICPS systems of the future. A key problem within this project is identifying the best-fit industrial agent practices for a given ICPS. This paper reports on the design and development of a tool to address this challenge. This tool, called `IASelect`, is built using graph databases and provides the ability to flexibly and visually query a growing repository of industrial agent practices relevant to ICPS. `IASelect` includes a front-end that allows industry practitioners to interactively identify best-fit practices without having to write manual queries.*

## 1. INTRODUCTION

Industrial Cyber-physical systems (ICPS) are seen as a core ingredient in the 4th Industrial Revolution [1], which complemented with emergent ICT technologies, such as Internet of Things, Big Data, Cloud Computing and Data Analytics, promotes the deployment of more interoperable, flexible, responsive and reconfigurable devices and systems. ICPS contain deep integration of computational applications with physical automation devices and are designed as networks of interacting cyber and physical elements [1][2]–[3]. Each component of an ICPS integrates its physical hardware function with a software (cyber) application acting as a virtual representation of its tangible counterpart.

The Multi-Agent Systems paradigm, derived from distributed artificial intelligence, promotes distribution, decentralization, intelligence, autonomy and adaptation, contributing to achieve flexibility, robustness, responsiveness and re-configurability [4]. This paradigm provides a fundamentally different way to design complex control systems based on the distribution of intelligence and decentralization of control functions over distributed autonomous and cooperative entities, called agents. Used in industrial contexts, agents, or more specifically Industrial Agents, can help to develop highly adaptive ICPS. In industrial environments, and aligned with the ICPS principles, the interconnection of intelligent software agents with the automation control devices, e.g., robots and PLCs (Programmable Logic Controllers), assumes a crucial role. Usually, this interconnection is created in a proprietary, case-by-case, and ad hoc manner. However, the use of a standardized way to implement this interface can help achieve transparency, interoperability, and scalability.

The IEEE P2660.1 standardization project, "Recommended Practices on Industrial Agents: Integration of Software Agents and Low Level Automation Functions", has been working on a methodology to rank and select best-fit Industrial Agent practices for the interfacing between software agents and automation control devices. Previous work was devoted to identifying the patterns derived from a survey of existing implementations of industrial agents [5], and to assess their characteristics, using the ISO/IEC 25010 standards family as a starting point [6].

This paper describes the design and development of a tool called `IASelect` for implementing the methodology to select recommended interfacing practices. `IASelect` uses a graph database to store interfacing practices templates and their technological instantiations along with their characterization according to a set of quality criteria.

It provides a front-end for users to interactively retrieve the best interface practice for a particular application scenario. In particular, the primary contributions of this paper are:

1. The design and development of a graph database to store the available data on Industrial Agent practices. This approach provides several benefits including, better data governance, data visualization, and interactive querying. A summary of available industrial agent practices is discussed in Sec. 2 and the design of the graph database is presented in Sec. 3.

2. The creation of query patterns and templates to allow industrial practitioners to use and query graph databases more easily. These patterns allow more flexibility than more static mechanisms like forms and spreadsheets. We discuss the design and development of these patterns in Sec 4.

3. An implementation of the proposed graph database and query patterns using Neo4j and Java into a tangible tool called `IASelect`. This GUI-based tool can be used by both users and administrators to identify practices and/or manage the knowledge base. The implementation is presented in Sec. 4

## 2. BACKGROUND

Software agents can work with low-level automation functions in a variety of ways. A survey of commonly-encountered practices helped the P2660.1 working group to develop a set of generic interface practices clustered according to two dimensions, as illustrated in Fig. 1 [7]. *Coupling*, shown on the X-axis, is dependent on the integration between high-level control (agents) and low-level control. Tight coupling indicates a direct and permanent coupling, as in the use of remote procedure calls. Loose coupling involves a mediated connection, such as through a queue. The Y-axis pertains to the *location* of the agent. Agents can be *on-device*, where they run on the same controller as the low-level functions. *Hybrid* systems have agents running externally rather than on the same controller. The survey carried out by the working group classified available practices into four primary interface practices: *Tightly Coupled–Hybrid*, *Tightly Coupled–On-device*, *Loosely Coupled–Hybrid* and *Loosely Coupled–On-device*. Each one of the generic interfacing practices shown in Fig. 1 can be instantiated using several different technologies.

Each practice has an associated set of qualities or characteristics, which make it more suitable for use in specific contexts. Selecting a best-fit interface practice for a given system context, therefore, requires identifying these associated qualities for each practice. The P2660.1 working group used the comprehensive yet generic set of characteristics from ISO/IEC 25010 [8], formerly ISO/IEC 9126, as a starting point to differentiate between practices. The ISO/IEC 25010 standard groups system qualities into eight characteristics. Each characteristic is then further separated into multiple sub-characteristics. Subsequently, a survey was conducted with a team of experts in the domain to identify qualities that are most relevant to the field [6]. The survey found that *testability, availability, time*

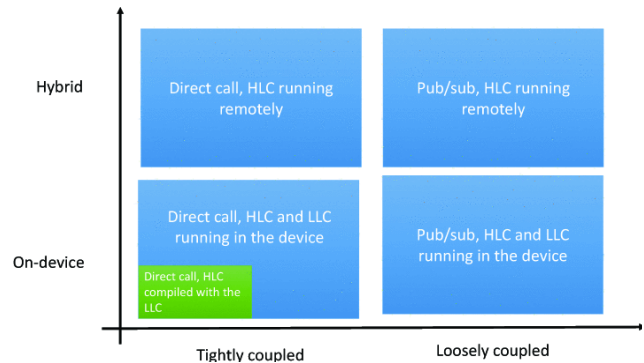

Fig. 1. Interface patterns considering interaction mode and location levels of abstraction. [7]

*behavior, interoperability, availability, fault-tolerance* and *reusability* emerged as the most important characteristics for the practices. Further work conducted by the working group showed how specific measures from the standard could be used to evaluate the practices [9]. In this paper, the characterization of practices and implementations is extended into a tool that allows stakeholders, especially industry experts, to identify best-fit practices through the qualities that are most desirable in their context.

There are existing tools used for managing and querying data sets for domains similar to ICPS. For instance, in [10] authors discuss an approach to store information related to security standards in relational databases and Structured Query Language (SQL) is used for data retrieval. In [11], authors have discussed an approach to store security requirements in a schema less XML database. However, a database with no schema based restrictions has higher risk of data corruption. In [12], authors have presented a tool to visualize requirements, and Neo4j database has been used to maintain the graph structure. Similar kind of tools have also been used in other domains for example in [13] where authors have proposed a tool for storing chemical compounds as graphs and graph algorithms are used to search chemical structures over the database. In [14], authors have proposed an approach based on graph databases for genome sequencing. The approaches discussed so far, except [10] use graph theory concepts to handle and inquiry data. In this paper, we present a tool that uses a graph database to store the P2660.1 data set and, automates the analysis of existing interfacing practices on user-defined selection criteria.

## 3. DESIGNING A GRAPH DATABASE FOR SELECTING INDUSTRIAL AGENT PRACTICES

As data size increases, managing data with traditional tools such as spreadsheets becomes a complicated task. Based on the law of entropy, an increase in data size also means that over time, disorder in the data set will increase. Spreadsheets are too cumbersome to maintain, primarily when shared and used by multiple stakeholders, such as users and administrators simultaneously. Database management systems serve as an alternative for organizing large data sets. Furthermore, they assist in correlating and analyzing collected data.

TABLE I. SYSTEM QUALITIES MAPPED IN THE P2660.1 DATA SET

| Domain | Function | Maintenance | Performance Efficiency |
|---|---|---|---|
| Factory Automation Building Automation Energy | Monitoring Control Simulation | Re-usability Capacity To Host agents | Time behaviour Scalability |

### A. Rationale for using Graph Databases

Rational database management systems (RDBMS) are the most popular tools for managing data. They have proven to be persistent in providing concurrency control and integration mechanism for data since 1970s [15]. RDBMS are highly efficient in handling large data banks. However, RDBMS have limited ability to capture the overall semantics of a domain [15], [16]. Moreover, as the number of relationships between data grows, RDBMS become inefficient in managing and querying data [17]. On the other hand, Graph databases (GDBs) are gaining wide acceptance in the industry due to there application in domains that deal with the querying and analysis of connected data [15], [17]–[21]. A graph database contains nodes and edges where nodes represent the entities and edges represent relationships between the entities [17], [18]. Together, nodes and edges capture the overall semantics of the domain. The resulting structure is more straightforward and is at the same time more expressive than those produced by RDBMS and Not Only SQL (NO-SQL) databases such as wide-column stores, document stores and key-value stores [22], [23].

For searching data, a spreadsheet or a RDBMS performs a search and match operation. This operation represents a relational join between different tables to calculate relationships at the time when a query is running. This operation tends to be computationally expensive in highly interconnected datasets. Graph databases are more efficient in such cases as the relationships between data are created at database creation stage and are stored inside the database. Hence, the overhead of calculating relationships at the time when data is being retrieved from the database is minimized in graph databases.

Graph database solutions such as Neo4j are based on the property graph data model [15], [19]. A property graph data model is more expressive than other graph data models, such as the resource description framework (RDF) [24]. A property graph stores information inside nodes and edges as key-value pairs which means that information can be embedded inside relationships which is an advantage over the RDF data model.

Another advantage of using graph databases is that they scale well. A property graph data model supports multigraphs where two nodes can be connected via multiple edges with each edge containing separate information about the relationship between the two nodes. Adding more information, therefore, does not require a refactoring or restructuring of the database. Current graph database solutions such as Neo4j are schema optional [24], which means that the graph database can easily accommodate any structural changes. While there are higher chances of data corruption, this risk can be mitigated by enforcing integrity constraints and writing additional logic in a programming language like Java or Python. The use of such integrity constraints at the database creation stage ensures data integrity and data consistency.

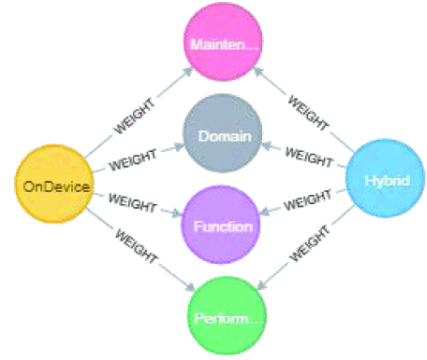

Fig. 2. Graph Schema for storing information about practices

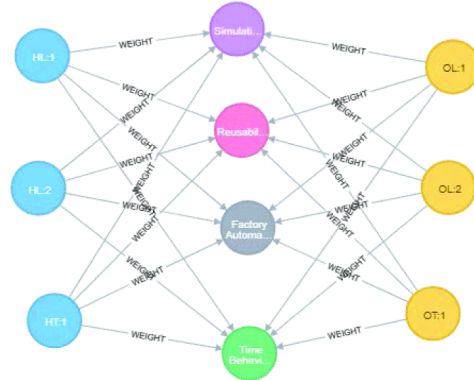

Fig. 3. Sample Graph Database representation of P2660.1 data set

### B. A Graph Database for Industrial Agent practices

The P2660.1 data set relates each interfacing practice to a score for specific system qualities, some of which come from ISO/IEC 25010. Table I shows the list of these qualities. This mapping was represented as a two-dimensional adjacency matrix where each mapping between a practice and a sub-characteristic was assigned a value, called its *weight*. The adjacency matrix can be visualized as a graph where interfacing practice and sub-characteristics are represented as nodes. Furthermore, the edge between a practice and a sub-characteristic is labeled with the appropriate score for that relationship.

*1) Graph Schema for P2660.1 data set:* Creating a graph database requires information about how data can be connected and structure. This information is called a *graph schema*. A graph schema provides an general view of the entire database by capturing its topology. Intuitively, nodes and edges of the graph schema represent the node, and edge types of the graph database. Node and edge types also assist in grouping together the nodes and edges of the graph database later for searching and visualization.

Fig. 2 shows the graph schema constructed using the P2660.1 data set. The schema is a labeled directed graph where the node types represent the relevant characteristics from Table I, as well as the two possible location levels (OnDevice and Hybrid). The graph schema allows storing the weights on edges that reflect the scores as per the P2660.1 data set. The direction of an edges shows that there exists a weighted mapping from a practice to a characteristic. For example, in the graph schema in Fig. 2

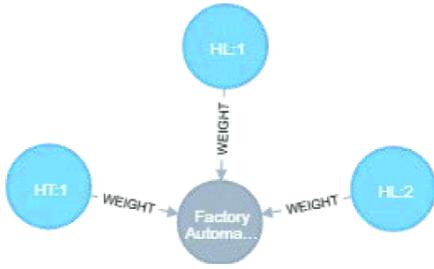

Fig. 4. Result of running Query [1] on the Graph Database shown in Figure 3

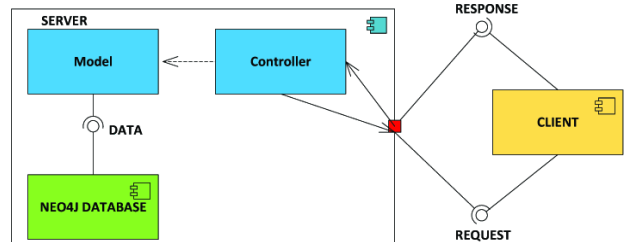

Fig. 5. Component Diagram for `IASelect`

there is an outgoing edge from hybrid practice and an incoming edge to maintenance.

*2) A Graph Database for the P2660.1 data set:* A graph database (GDB) is instantiated from its graph schema. Fig. 3 represents the graph database for the P2660.1 data set, instantiated from the graph schema in Fig. 2. Information related to nodes and edges of the graph database is contained as attributes of these elements and are stored as key-value pairs. For example, the node with name OT : 1 represents a OnDevice tightly technique where the attributes key apiClient has a value java assigned to it. Similarly, there are edges between nodes where attribute key weight has a value assigned to it.

**C. Querying the P2660.1 Graph Database**

In this research, we are using Neo4j as a graph database for storing data. To retrieve data from the database we need to define the kind of data we want to extract. Neo4j provides a declarative query language called Cypher [24] to retrieve data from the graph database. Searching a graph database requires defining a sub-graph as a pattern to look for within the database. This sub-graph is expressed using a *pattern graph*.

A pattern graph assists in defining the sub-graph of interest so that similarly structured data can be extracted from the database. Structurally, a pattern graph is expressed as a sub-graph of the graph schema. A pattern matching algorithm then uses the pattern graph to search for the sub-graph over the graph database. For example, by referring to the graph schema as in Fig. 2 one can search for queries such as find all hybrid techniques for factory automation domain which have been assigned weight greater than 2. Such a query can be expressed as a pattern graph in Cypher as follows:

```
MATCH(h:Hybrid)-[w:WEIGHT]->(d:Domain)
WHERE w.value > 2
AND d.name = "Factory Automation"
RETURN *                              [1]
```

The pattern graph is expressed in the MATCH clause of the query. A pattern graph consists of node/edge types and node/edge variables. The node/edge types assist in specifying the type of data, and the node/edge variables assist in accessing the node/edge attributes of the graph database. The MATCH clause uses a pattern matching algorithm to find all the matching sub-graphs in the graph database. The sub-graphs are further restricted based on the filter conditions specified in the WHERE clause. Filter conditions are set based on the attribute value stored in the database and are designed using the node/edge variables. Finally, the RETURN clause outputs the sub-graph shown in Fig. 4 and marks the end of the query.

## 4. IMPLEMENTATION

Graph databases are at an early stage of industry-wide adoption. Moreover, users working in domains such as cyber-physical systems and multi-agent systems may not be familiar with graph database query languages like Cypher. Therefore, we have developed a tool `IASelect` that assists in querying graph databases without requiring a working knowledge of Cypher.

**A. `IASelect`-Architecture**

IASelect must feature several important qualities. We use the terminology from ISO/IEC 25010 to list the following system characteristics:

- *Functional suitability:* Functionally, `IASelect` must provide features such as the ability for administrators to manage the underlying database, and the ability for users to query the database to rank available practices that are relevant to their context.
- *Usability:* `IASelect` must be highly usable for both administrators and users. It must allow users to enter information interactively and provide appropriate user error protection, and also present the results clearly. The tool must be accessible for multiple users from different sub-domains of industrial control.
- *Availability:* `IASelect` must be accessible to multiple users, possibly present in different locations, at the same time.
- *Portability:* `IASelect` should be independent of the users' computer configurations.

*Functional suitability* is supported through the design of the database, as described in Sec 3, which allows all desired features to be included within the tool. The architecture of `IASelect`, shown in Fig. 5, supports all other characteristics.

To achieve high *availability*, `IASelect` is based on a client-server architecture. The client side is a web-page that can be run on most machines (supporting *portability*). The server runs both the Neo4j database and an application to handle requests from the client. Decoupling the server side from the client's machine allows us to (a) allows users to use `IASelect` without installing any new software such as Neo4j, (b) control the server side for both privacy and performance, and (c) allow for easily modifying or scaling the server or the client side without affecting the other. The application server provides a restful web service so that users can query graph database over the web. Currently, the server application and database are deployed on a cloud data center.

For *usability*, which is a primary characteristic of `IASelect`, we embed a Model-View-Controller (MVC) design pattern inside the client-server architecture. The client-side web-page contains the View which can change depending on who the user is. Currently, we support two views: the administrator view and user view. Administrators can to update the database while users can only query it. At the time of writing, only the user view has been fully integrated into `IASelect`. The Model and Controller are java class objects which run on the server side application. The Controller class object handles the conversion of requests from a View into Cypher queries, and the Model class runs the queries on the Neo4j database. Using the MVC pattern support scalability and enables the addition of Views (for additional user types) easily. It also ensures that the code base is cleaner and understandable, making it easier to maintain.

The client web-page further improves *usability* by providing drop-down lists (for reducing user error) for users to select appropriate context-specific qualities and metrics. The results returned from the server side application are then displayed as a ranked list which can easily be understood.

### B. Software Implementation of `IASelect`

*1) Technology Stack:* The server side has been implemented in Java and integrated with Gradle and Maven. Gradle is used for build automation. We have added the Spring boot plugin to the Gradle project to provide an embedded Tomcat server to host the server side application. We use the Maven libraries to connect the Java project with the Neo4j database. The database queries written in Cypher are embedded inside the server side application's Java code and are executed through the appropriate method call. The client side is a web-page that is built using HTML5, CSS3, and Javascript. We have used AJAX to communicate between the client and the server using the XMLHttpRequest.

*2) Transaction Sequence:* The web-page running on the client machine is accessed through a URL. The tool assists users in extracting data from the graph database based on the specified criteria. `IASelect` generates a report which lists all the matching practices and recommends the most suitable practice. For generating a practice report in `IASelect`, users are presented with a web form. The web form serves as a boilerplate [25] to specify the criteria for generating a practice report. Boilerplates are semi-complete query structures that can be completed through user input.

When a request is submitted, the controller object running within the server side application receives the request. At the same time, the application establishes a connection with the Neo4j database instance using the Bolt protocol. Bolt is a TCP based network protocol which is integrated into Neo4j for connecting to other applications. Once the connection between the database and the model has been established, the controller passes the request to the model by calling the appropriate method. The model then requests a session with the Neo4j database instance and sends a query written in Cypher to be executed at the Neo4j database. The request parameters received from the client are embedded inside the Cypher query. The model then returns the query results obtained from Neo4j database to the controller. The controller then passes the result-set to the server and, finally the server sends the result-set back to the client in the form of a response. At the client, the response is further processed and is displayed on the web page.

*3) Tool Usage:* The user provides the necessary context-specific details using the following steps. In the first step, a user sets the context of search by specifying relevant qualitative requirements that the Industrial practices must fulfill. This is done by selecting the sub-characteristics related to function and domain as listed in Tab. I. The sub-characteristics defines the application context for the interface practice. Furthermore, the user also specifies if the practice should be capable to host agents. For example, as shown in section 1 of Fig. 6, the context is set for searching practices for factory automation domain, simulation function and the practice should be capable to host agents. A practice report can be generated for other sub-characteristics of function and domain by using the drop down menu in the web form.

In the second step, the user sets criteria based on maintenance and performance efficiency related to the practices. Users specify which sub-characteristics are deemed most relevant in their context. For determining the relevance of sub-characteristics a percentage scale is assigned on the weights between practices and sub-characteristics related to maintenance and performance efficiency. For example, as shown in section 2 of Fig. 6, scalability, time-behaviour and re-usability are set with percentage scale of 10, 10 and 80 respectively (the total must be 100). In this particular scenario, the user clearly prefers a practice with high level of re-usability, and that scalability and time-behaviour are of lower relevance.

Finally, based on the context and criteria, a practice report is generated which displays a list of matching practices with technique name, API client, channel and final score assigned to each practice. The final score is calculated by multiplying cumulative percentage weight for each practice with a respective average weight between practice and particular function sub-characteristics. For example, as shown in section 4 of Fig. 6 technique *HL:2* has *Apache Milo*, *MQTT* and, *4.6* as API client, channel and, final score, respectively. The recommended practice as highlighted in section 5 of Fig. 6, corresponds to the practice that got the highest final score, for example, *HL:1* is the recommended practice for this scenario. The tool also provides the list of alternative interface practices sorted based on the score values.

## 5. CONCLUSIONS AND FUTURE DIRECTIONS

This paper presents the construction of a graph database tool, called `IASelect`, to allow industry practitioners to identify best-fit industrial agent practices for industrial CPS. `IASelect` is easy to use and, its architecture enables scalability and flexibility. For instance, the edges of the graph database currently contain the weights between practices and sub-characteristics, which makes the database equivalent to a spreadsheet table. In the future, additional properties can be added to nodes and edges without altering the topology of the graph. Such scalability and flexibility are not present in spreadsheets.

Fig. 6. Web form based client interface of IASelect

The front end of IASelect enables users to query graph databases without having a working knowledge of query languages like Cypher. IASelect uses a boilerplates based approach that enables users to query the graph database. Furthermore, the boilerplate based approach is not limited to the P2660.1 data-set and can be extended to domains other than ICPS. IASelect has been deployed in the cloud as a restful web service. This enables other users to access data related to industrial agent practices via Restful web API. Furthermore, users can integrate the web service into their own applications. Deploying IASelect in the cloud also provides advantages specific to cloud computing technology such as scalability, availability, reliability and security.

IASelect is an attempt to harness the potential of property graph databases in the domains such as ICPS. However, currently we are only partially utilizing the power of graph database query languages. Graph databases enable users to identify, search and, extract patterns from data. Users can specify a sub-graph of interest to search all similar occurrences of sub-graph in graph database. In IASelect however, we are searching for very specific patterns which are tailor made to meet the requirements from the P2660.1 standard and it cannot yet be used to search for generic patterns. In the future, IASelect also needs to feature an administrator view for inserting new data, and updating and deleting data from the database.